\documentclass{appolb}
\usepackage{graphicx}
\usepackage{cite}
\usepackage{csquotes}

\begin{document}
\eqsec  
\title{`Phase Transition' in the `Thorium-Isomer Story' %
\thanks{Presented at the XXXVI Mazurian Lakes Conference on Physics, Piaski, Poland, September 1-7, 2019.}%
}

\author{P.G. Thirolf$^{a}$, B. Seiferle$^{a}$, L. v.d. Wense$^{a}$, I. Amersdorffer$^{a}$,
        D. Moritz$^{a}$, J. Weitenberg$^{b}$,
\address{$^a$ Ludwig-Maximilians-Universit\"at M\"unchen, Garching, Germany \\ 
         $^b$ Rheinisch-Westf\"alische Technische Hochschule, Aachen, Germany}}  

\maketitle
\begin{abstract}
 Given the drastic progress achieved during recent years in our knowledge on the
 decay and nuclear properties of the thorium isomer $^{229m}$Th, the focus
 of research on this potential nuclear clock transition will turn in the near
 future from the nuclear physics driven `search and characterization phase'
 towards a laser physics driven `consolidation and realization phase'. This prepares
 the path towards the ultimate goal of the realization of a nuclear frequency standard, the 
 `Nuclear Clock'. This article briefly summarizes our present knowledge, focusing on recent
 achievements, and points to the next steps envisaged on the way towards the Nuclear Clock. 
\end{abstract}
  
\section{Introduction}

\begin{figure}[htb]
  \centerline{
  \includegraphics[width=8.cm]{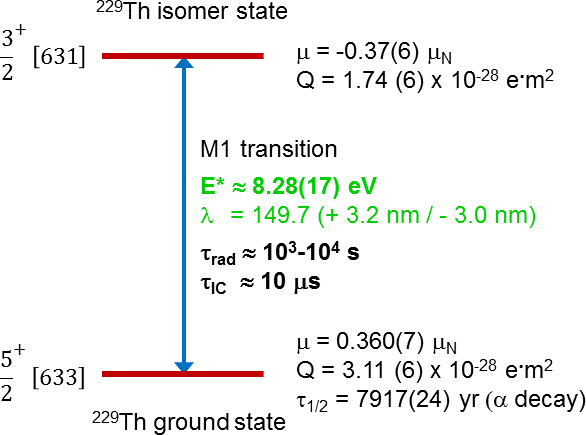}}
  \caption{Properties of the $^{229}$Th ground and isomeric first excited state.
           Nuclear levels are given with their spin, parity and Nilsson classification.
           $\mu$: magnetic moment, Q: electric quadrupole moment.}
  \label{Fig1:2-level-properties}
\end{figure}

The `thorium isomer' in $^{229}$Th denotes the first excited state in $^{229}$Th, an exotic 
singularity in the landscape of presently more than 3300 known isotopes with their currently more 
than 184000 excited nuclear states. The thorium isomer $^{229m}$Th exhibits the lowest nuclear excitation 
in all presently known nuclides, 4 to 5 orders of magnitude lower than typical nuclear excitation 
energies. Fig.~\ref{Fig1:2-level-properties} summarizes our present knowledge of the nuclear 
properties of the $^{229}$Th ground and isomeric first excited state.

Early on, the wide application potential arising from the unique properties of the thorium 
isomer for various physics disciplines like \enquote{optics, solid state physics, lasers, plasma physics, 
and others}~\cite{strizhov1991} was realized. Amongst them frequency metrology plays a central role, 
as the thorium isomer is placed in the energy and half-life region of atomic transitions 
exploited for high-precision optical atomic clocks. Its small relative natural linewidth 
of $\Delta$E/E$\approx$ 10$^{-20}$ (derived from the expected long half-life of a few 10$^3$ 
seconds~\cite{tkalya2015}) and the about 5 orders of magnitude smaller nuclear electromagnetic 
moments compared to those of atoms render the thorium isomer a natural candidate for a highly 
precise nuclear frequency standard, a nuclear clock, which was first proposed by Peik and Tamm 
in 2003~\cite{peik2003}.  

Starting from its first indirect conjecture from $\gamma$ spectroscopic data in 
1976~\cite{krogerreich1976}, intensive experimental efforts were dedicated towards a direct 
detection of the thorium isomer and improved knowledge of its properties, in order
to be able to exploit its unique potential for applications as a nuclear frequency standard,
in geodesy or for fundamental studies like the identification of potential variations of
fundamental constants like the fine structure constant. Details can be found in the review 
articles~\cite{wense2018,thirolf2019a,thirolf2019b}.

\section{Current knowledge on the properties of the nuclear clock 
           isomer $^{229m}$Th}

\begin{figure}[htb]
  \centerline{
  \includegraphics[width=12.5cm]{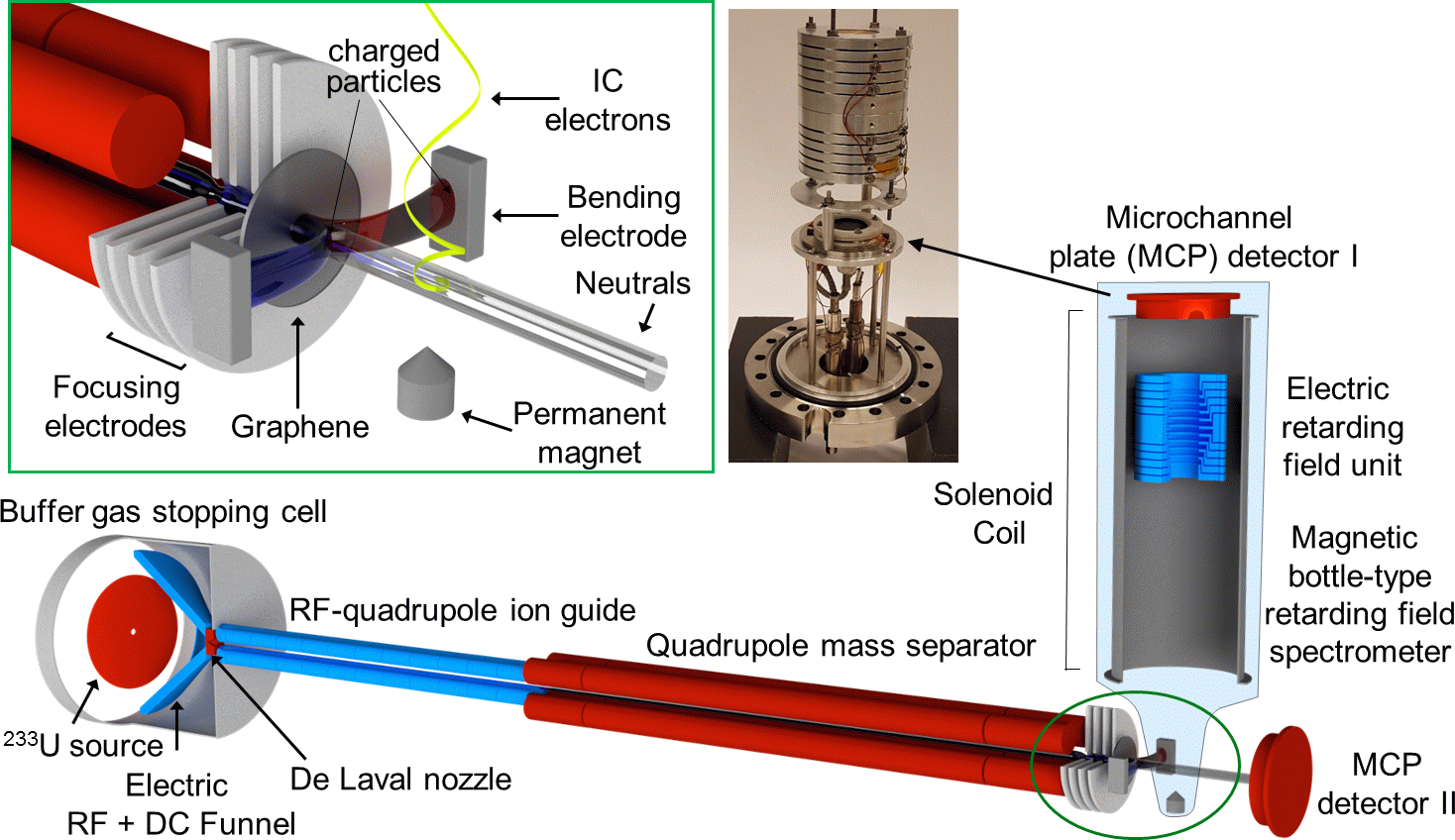}}
  \caption{Experimental setup for the determination of the thorium isomer's
           excitation energy, complementing the buffer-gas cell based `isomer generator' 
           with a magnetic bottle electron spectrometer (retarding field unit 
           shown as photograph in the inset).
           Adapted from Ref.~\cite{seiferle2019}.}
  \label{Fig3:pic-espectrometer}
\end{figure}

The last few years have seen several major experimental breakthroughs in our understanding 
of the exotic thorium isomer, starting with the first direct detection of the isomer's
ground state decay in the internal conversion (IC) decay channel published in 2016~\cite{wense2016},
the first determination of the lifetime of the neutral thorium isomer in 2017~\cite{seiferle2017},
the first determination of the hyperfine structure of $^{229m}$Th via collinear laser 
spectroscopy~\cite{thielking2018} to the most recent milestone of the first direct
measurement of the thorium isomer's excitation energy, again via the IC decay 
channel~\cite{seiferle2019}. All of these results were based on the production of an 
isotopically pure beam of $^{229(m)}$Th ions (from the $\alpha$ decay of a $^{233}$U source),
extracted from a buffer gas cell and purified by a quadrupole mass separator~\cite{wense2019}.\\
Here, the most recent achievement of a direct decay energy measurement will be described
in more detail. The basic idea is to measure the kinetic energy of the electrons emitted
during the IC decay and to derive the excitation energy of $^{229m}$Th from there. 
In order to avoid the influence of surface effects, neutralization of the $^{229m}$Th 
ions followed by the IC decay (occurring within few microseconds by emitting an electron) 
was initiated by sending the ions through a bi-layer of thin graphene foils (set at -300 V). 
For these measurements the `standard' setup used to generate a $^{229m}$Th ion beam was 
complemented by a magnetic-bottle-type retarding field electron spectrometer. 
The experimental setup can be seen in Fig.~\ref{Fig3:pic-espectrometer}, in addition to 
the schematic overview a photograph of the home-built retarding field electron spectrometer 
can be seen in the inset.  

\begin{figure}[htb]
  \centerline{
  \includegraphics[width=9cm]{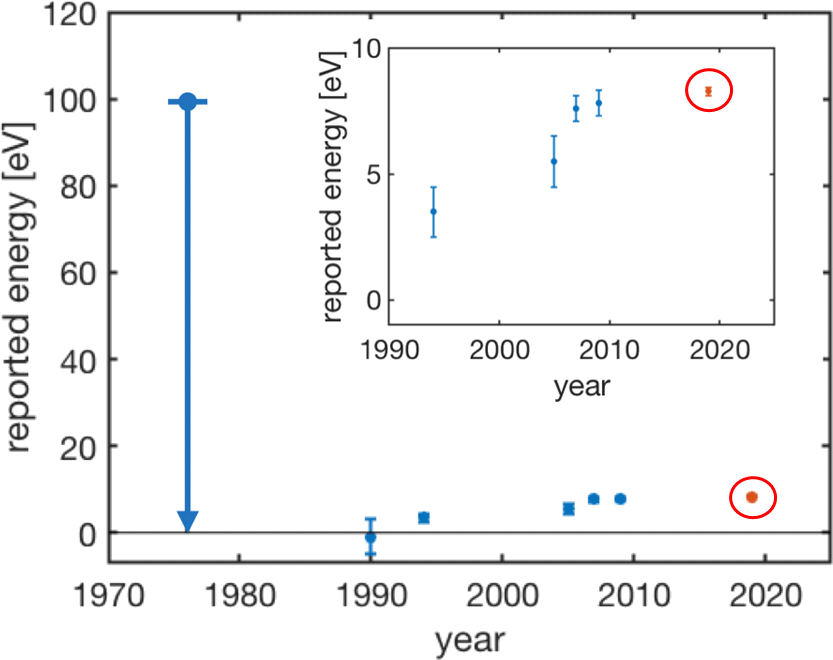}}
  \caption{Improvement of our knowledge on the excitation energy of the 
           thorium isomer $^{229m}$Th from its conjecture in 1976 to the recently
           determined most precise value from Ref.~\cite{seiferle2019} (encircled).}
  \label{Fig3:energy-history}
\end{figure}

The spectrometer~\cite{seiferle2019b} consisted of a strong permanent magnet that
generated an inhomogeneous magnetic field (about 200 mT in the area above the magnet) and
a solenoid coil creating a weak homogeneous transport field (typically 2 mT). 
Internal conversion electrons emitted in a spherical volume with a radius of about 1 mm 
above the permanent magnet were collected by the magnetic field and directed into the 
solenoidal transport field and guided towards a retarding-field unit placed in the 
solenoid coil. The kinetic energy of the IC  electrons can be analyzed by applying a retarding 
voltage to a gold grid (surrounded by ring electrodes to ensure a smooth gradient 
and terminated by additional gold grids) and by counting the electrons that reach the 
multichannel-plate electron detector (MCP I in Fig.~\ref{Fig3:pic-espectrometer}).
Only electrons with kinetic energies sufficient to overcome the applied retarding voltage 
were registered with MCP I.  
This resulted in an integrated spectrum that decreased 
monotonically with increasing retarding voltage. The spectrometer reached an excellent 
relative energy resolution $\Delta$E/E (FWHM) of about 3$\%$ (ca. 30 meV for 1 eV electrons).\\
For the determination of the thorium isomer's excitation energy it is most important to
notice that the neutralization process will typically end in excited electronic states E$_i$
of the neutral $^{229}$Th atom, while the subsequent IC decay leads to excited electronic 
states E$_f$ of the resulting thorium ion. Thus the kinetic energy of an IC electron, E$_{\rm kin}$, 
is connected to the energy of the isomer, E$_{\rm is}$, via 
\begin{equation}
   E_{\rm kin} = E_{\rm is} - IP + E{_i} - E{_f}
\end{equation}
where IP is the thorium ionization 
potential (6.308~$\pm$~0.003 eV~\cite{lias2005}). The transition rate for specific 
(E$_i$, E$_f$) pairs can be precisely predicted~\cite{pavlo-diss}.\\
Density functional theory (DFT) calculations allow to identify the relevant energy range
for E$_i$ (wavenumbers below 20000 cm$^{-1}$), where atomic theory identified 82 excited 
electronic states E$_i$ in thorium, each of them contributing (with unknown statistical 
population probability) to the deexcitation via IC to about four final states 
E$_f$~\cite{seiferle2019}. Given the non-observation of clear transition
lines in the integrated electron energy spectrum and the instrumental energy resolution of about 
30 meV, it could be concluded that at least 5 initial states must have contributed to the measured
electron spectrum. This number was taken as input for a statistical analysis based on simulated
energy spectra using random distributions of initial electronic states. The details of the method
that was used is comprehensively described in Ref.~\cite{seiferle2019}, the analysis resulted in
the presently most precise value for the excitation energy of the thorium isomer of 
8.28$\pm$0.17 eV (corresponding to a wavelength of 149.7$\pm$3.1 nm).\\
Fig.~\ref{Fig3:energy-history} gives an overview on the temporal development of our knowledge
of the $^{229m}$Th excitation energy from its conjecture in 1976 to the recently
determined most precise value (highlighted by circle).

Having achieved a first direct determination of the thorium isomer's ground-state transition
energy with 3-fold improved precision compared to the value adopted during the last decade
only marks an intermediate step towards the ultimate goal of realizing a nuclear clock, as will 
be further outlined in the next section.

\section{Perspectives towards a Nuclear Clock }

\begin{figure}[htb]
  \centerline{
  \includegraphics[width=12.5cm]{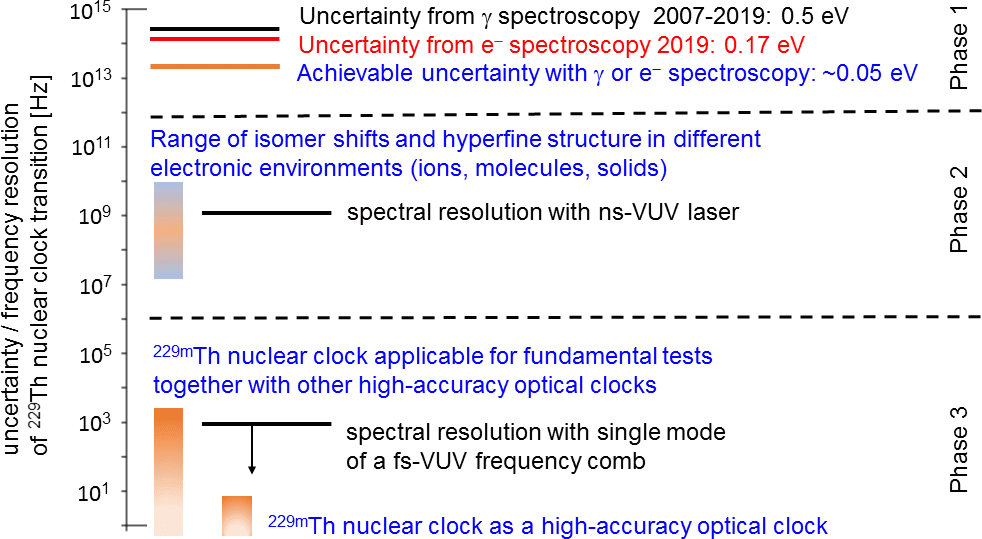}}
  \caption{Frequency scale of uncertainty and spectral resolution of the nuclear ground-state
           transition frequency from the thorium isomer $^{229m}$Th.}
  \label{Fig4:frequency-ladder}
\end{figure}

Fig.~\ref{Fig4:frequency-ladder} illustrates the challenges lying still ahead
on the way towards a high-precision nuclear clock: having recently improved the
precision of the $^{229m}$Th isomeric excitation energy still leaves us about
14 orders of magnitude away from the ultimate goal of an ultra-high precision
nuclear frequency standard on the Hertz level. However, a clear roadmap can be depicted that puts
this seemingly tremendous difference into a realistic scenario for realization 
within the next 5-6 years, staged in 3 phases. 

\begin{figure}[htb]
  \centerline{
  \includegraphics[width=10cm]{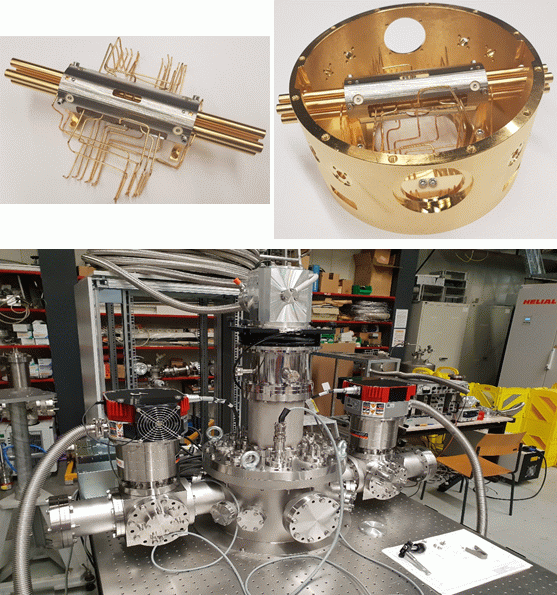}}
  \caption{Photographs of the cryogenic Paul trap under commissioning as the backbone for future
           laser manipulation (cooling, excitation) of $^{229m}$Th ions aiming at establishing a 
           nuclear frequency standard based on the thorium isomer. Top panels: cold mass of the
           cryotrap consisting of the four-rod trap electrode system and its electrical contact
           rod (left) and its assembly together with the surrounding Cu vessel. All parts have 
           been gold plated for surface quality improvement. Bottom: assembly of the cryotrap 
           mounted to an optical table, with the cold head on top, coupled to the cold mass
           via an ultra-low vibration interface. An optimized $^{229m}$Th injection system
           with a new gas cell, extraction RFQ and quadrupol mass separator (QMS) is shown on the
           left, while the extraction side on the right features another QMS and a detection section.}
  \label{Fig5:cryotrap-photos}
\end{figure}

After a period of more than 10 years
where no improvement on the isomeric excitation energy (7.8(5) eV~\cite{beck2007,beck2009} 
could be achieved, as presented before a first direct measurement with a precision 
improvement by about  a factor of three resulted in the currently best value of 
8.28(17) eV~\cite{seiferle2019}. In view of the dynamic activities presently dedicated to
research on $^{229m}$Th in many groups worldwide, it can be expected that a further 
improvement via $\gamma$ or electron spectroscopy can further reduce the uncertainty to
about 0.05 eV as the endpoint of what can be regarded as the still nuclear-physics dominated 
`phase 1' of experimental efforts.
This phase will also see extensive efforts to detect VUV photons following non-resonant 
nuclear excitation.
Moreover, the still unknown lifetime of the ionic isomer can be
experimentally targeted in a cryogenic Paul trap, which has been set up and is presently under 
commissioning in Garching. Conceptually based on the design of the CryPTex trap built at 
Max-Planck-Institute for Nuclear Physics in Heidelberg~\cite{schwarz2012}, long storage 
times in excess of 10 hours will be realized,
sufficiently long to cover the full expected range of 10$^3$ - 10$^4$~s. Fig.~\ref{Fig5:cryotrap-photos}
displays photographs documenting the present status of the Garching Paul trap setup. The top row
shows the four-rod trap electrode system and its electrical contact rods (left), while the right
panel shows the electrode system already mounted in the surrounding inner Cu vessel, together 
forming the cold mass of the cryogenic trap finally operating at liquid helium temperature. 
All parts have been gold plated to achieve optimum surface quality. The bottom picture shows
the full assembly of the cryotrap, which is mounted to an active vibration-compensated optical table.
The interaction chamber sits on top of a circular hole in the optical table, such that the main
turbomolecular vacuum pump of the setup can be mounted from underneath the table. On top of the
main vacuum chamber with the cold mass and 12 in-plane and 4 out-of-plane access ports for
ion injection and extraction, diagnostics and optical manipulation a pulse tube cold head 
is mounted, vibrationally decoupled by an ultra-low vibration interface.

On the left side of the main chamber the injection section will be mounted, comprised of a new
buffer gas stopping cell (housing the $^{233}$U recoil source, an RF/DC funnel and a de-Laval 
extraction nozzle) followed by an extraction radiofrequency quadrupole (RFQ) and a subsequent
quadrupole mass separator (QMS). On the extraction side another QMS and a detection section
is mounted. Pumping is provided by turbomolecular pumps at the injection and extraction side and 
underneath the main chamber. The cryotrap setup is presently being finalized and commissioning tests
will start in late 2019. Subsequently, sympathetic laser cooling of Th$^{3+}$ ions with Sr$^+$ 
will be added to the trap setup (featuring a favorable mass-to-charge ratio and offering a suitable 
electronic level scheme for laser cooling down to mK temperatures). 
First measurements on stored and cooled $^{229m}$Th ions (1-10 ions simultaneously trapped) 
will target the ionic lifetime of the isomer.\\
From then on laser spectroscopy will dominate the methodology towards improved insight into
the properties of the thorium isomer. The biggest challenge and ultimate doorway to the realization
of a nuclear clock will be given by achieving a resonant laser excitation of the nuclear clock transition
in a second phase of experimental campaigns, based on already existing laser technology~\cite{wense2017}.
Nanosecond VUV laser systems with GHz frequency bandwidth will allow for a systematic search for the
nuclear resonance, while as well giving access to the spectral range of isomer shifts and hyperfine
structure in different electronic environments of ions, molecules and solids. \\
A third phase will start upon realization of resonant laser excitation of the thorium isomer.
Customized laser systems in the wavelength range around 150 nm will have to be developed, with a 
spectral bandwidth of 1 kHz and better, in order to allow for establishing a nuclear clock already 
able to contribute competitively to fundamental physics tests like variations of fundamental constants. 
With a laser system operating around 150 nm with a spectral bandwidth of 10$^2$-10$^3$ Hz (corresponding 
to ${\Delta f \over f} = 5\cdot 10^{-13} - 5\cdot 10^{-14}$), together with an expected sensitivity 
enhancement of about 10$^4$ provided by $^{229m}$Th for studies of temporal variations of the fine 
structure constant $\alpha$, better limits than the presently best value of 
${\dot{\alpha}\over\alpha} = (-0.7 \pm 2.1)\cdot 10^{-17} yr^{-1}$~\cite{godun2014} can be achieved.\\

\begin{figure}[htb]
  \centerline{
  \includegraphics[width=12.cm]{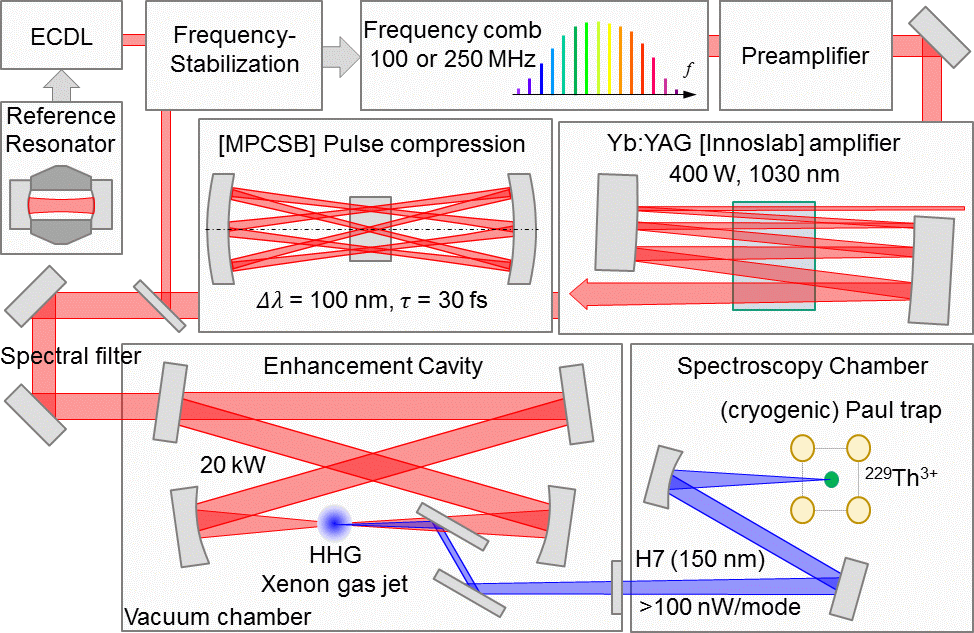}}
  \caption{Schematical outline of the VUV frequency-comb based laser designed
           to drive the nuclear clock transition in $^{229m}$Th.}
  \label{Fig6:laser-concept}
\end{figure}

Having learned from the recently refined value of the isomeric excitation energy that at least
currently no cw laser will be able to reach into the energy region of the thorium isomer's ground-state
transition around 150~nm, VUV frequency combs will be the present technology of choice to realize a laser 
system that will allow for operating a thorium nuclear clock.
Identification of a resonant excitation of $^{229m}$Th in the Paul trap will be achieved via a 
double-resonance technique as outlined in Ref.~\cite{peik2003}.
Fig.~\ref{Fig6:laser-concept} features a conceptual overview of such a high mode power VUV frequency 
comb system, based on the 7$^{\rm th}$ harmonic of 1030~nm generated in a Xe gas jet.

In this preliminary layout and specification, the laser is based on an ultra-low noise IR frequency comb 
(mode spacing 100 MHz or 250 MHz). IR frequency combs are available with high longterm stability, 
5$\cdot$ 10$^{-19}$ in 1000 s and 5$\cdot$10$^{-18}$ in 1 s could already be demonstrated~\cite{menlo}. 
The comb will be equipped with a near-IR (NIR) high-power output port at 1030 nm. 
The system will be stabilized to an external cavity laser (1542.14 nm), locked 
to an ultrastable high-finesse optical reference cavity. The comb output will subsequently be amplified 
to 400 W via a preamplifier and an Yb amplifier on the basis of the InnoSlab amplification 
technology~\cite{russbueldt2015}, providing a large amplification and high average power at high 
efficiency, while no chirped pulse amplification (CPA) is needed at multi 10 MHz repetition rate.
The output pulse exhibits a pulse length of about 700 fs with a 
pulse repetition rate of about 50 MHz. The amplification stage is followed by a 2-stage pulse 
compression, resulting in an output bandwidth of about 100 nm and a pulse duration of 30 fs. 
Finally, the compressed laser pulse will be injected into an enhancement cavity, where the 7$^{\rm th}$
harmonic of the driving IR (comb) radiation in a Xe gas jet will be generated. A laser power of $\geq$~1 nW 
per mode is conservatively 
targeted, while ultimately aiming for ($\geq$)100 nW/mode. The bandwidth will be $\leq$ 500 Hz, 
ultimately aiming at the few Hz regime. \\
Such a laser system could open the door to the operation of a nuclear clock and to experimental campaigns
addressing the unique applications specifically envisaged for fundamental tests of physics beyond the
Standard Model like time variations of fundamental constants.\\
With the presently ongoing dynamic experimental as well as theoretical efforts, stimulated by the 
recent experimental breakthroughs, the `phase transition' from the nuclear-physics driven identification 
and characterization phase to a laser-physics driven consolidation, realization and application phase 
in the more than 40 year old story around the thorium isomer and its potential as a nuclear frequency
standard is already ongoing. Further intriguing results originating from this exotic solitair in 
the nuclear physics landscape can be expected in the coming years.

\section*{Acknowledgments}
This work was supported by the European Union’s Horizon 2020 research and
innovation programme under grant agreement number 664732 (nuClock) and Grant Agreement 856415 
(ThoriumNuclearClock), by DFG (Th956/3-2) and by the LMU Chair of Medical Physics via the 
Maier-Leibnitz Laboratory Garching. 
We gratefully acknowledge fruitful discussions with J. Crespo Lopez-Urrutia and
T. Schumm.

\end{document}